\documentclass[12pt]{article} \def\theequation{\arabic{section}.\arabic{equation}}
\newcounter{rown}

\usepackage[centertags]{amsmath}
\usepackage{color}
\usepackage{amsfonts}
\usepackage{amssymb}
\usepackage{amsthm}
\usepackage{newlfont}
\begin{document}
\renewcommand{\theequation}{\arabic{section}.\arabic{equation}}
\title{{\bf Supertwistors, massive superparticles and $\kappa$-symmetry}}
\author{ J.A. de Azc\'{a}rraga, \\
Departamento de F\'{\i}sica Te\'orica and IFIC (CSIC-UVEG), \\
 46100-Burjassot (Valencia), Spain\\
 J. M Izquierdo,\\
Departamento de F\'{\i}sica Te\'orica, Universidad de Valladolid, \\
47011-Valladolid, Spain,\\
and \\
 J. Lukierski \\
Institute of Theoretical Physics, Wroc{\l}aw University, \\
50-204 Wroc{\l}aw, Poland}
\date{\small{August 15, 2008; revised November 28}}
\maketitle
\begin{abstract}
We consider a $D=4$ two-twistor lagrangian for a massive particle
that incorporates the mass-shell condition in an algebraic way,
and extend it to a two-supertwistor model with $N=2$ supersymmetry
and central charge identified with the mass. In the purely
supertwistorial picture the two $D=4$ supertwistors are coupled
through a Wess-Zumino term in their fermionic sector. We
demonstrate how the $\kappa$-gauge symmetry appears in the purely
supertwistorial formulation and reduces by half the fermionic
degrees of freedom of the two supertwistors; a formulation of the
model in terms of $\kappa$-invariant degrees of freedom is also
obtained. We show that the $\kappa$-invariant supertwistor
coordinates can be obtained by dimensional
($D$=6$\rightarrow$$D$=4) reduction from a $D=6$ supertwistor. We
derive as well by 6$\rightarrow$4 reduction the $N=2$, $D$=4
massive superparticle model with Wess-Zumino term introduced in
1982. Finally, we comment on general superparticle models
constructed with more than two supertwistors.
\end{abstract}
\newpage

\section{Introduction}

The conformal structure of twistor theory (see {\it e.g.}
\cite{1azluk}-\cite{4azluk}) implies that relativistic particles
described by single twistors are massless
\cite{1azluk,3azluk,5azluk,6azluk}. To describe massive particles
at least two twistors are needed
\cite{Perjes,7azluk,3azluk},\cite{8azluk}-\cite{12azluk} ({\it
cf.} \cite{BaPi}). Indeed, with two $D=4$ twistors
\begin{equation}
\label{eq1.1}
 Z_{Ai} = (\lambda_{\alpha i}\ ,\ {\bar{\omega}}^{\dot{\alpha}}_i)
  \ ,\quad i=1,2 \; ,\quad \alpha,\dot{\alpha}=1,2\;,
\end{equation}
the standard formula for the composite fourmomentum\footnote{
$\lambda^{\alpha i} = \epsilon^{ij} \lambda^{\alpha }_{\ j} $;
$\epsilon^{ij} = \left(
\begin{array}{cc} 0 &1 \cr -1 & 0
\end{array}  \right)$ ,
$\epsilon_{ij} \epsilon^{jk} = \delta^k_j$. In general,
$a_{\alpha{\dot\beta}}\equiv
\frac{1}{\sqrt{2}}\sigma^\mu_{\alpha{\dot\beta}}a_\mu\,$,
$\,\sigma_{\alpha {\dot \beta}}=(\sigma^0,\sigma^i)\;,\;
\sigma^{{\dot\beta}\alpha}=(\sigma^0, -\sigma^i)\,,\,i=1,2,3$.
Then, $a^\nu=\frac{1}{\sqrt{2}} a_{\alpha{\dot \beta}}\sigma^{\nu\
{\dot\beta}\alpha}\,$, $\,a_{\alpha{\dot \beta}}a^{{\dot
\beta}\gamma}={\textstyle {\frac{1}{2}}}\delta^\alpha_\gamma
\,a^\mu a_\mu \,$ and $a_{\alpha{\dot
\beta}}b^{{\dot\beta}\alpha}=a\cdot b$, $a^2=({a^0})^2-{\bar
a}^2.$ Complex conjugated spinors are denoted by $(\lambda_{\alpha
i})^*\equiv \bar{\lambda}_{\dot{\alpha}i}$, etc.}
\begin{equation}
\label{eq1.2}
 p_{\alpha \dot{\beta}}= \lambda_{\alpha  i} \; \bar{\lambda}_{\dot{\beta} i}
\quad , \quad p_{\mu}= {\textstyle \frac{1}{\sqrt{2}}}  p_{\alpha
\dot{\beta}}(\sigma_\mu )^{ \dot{\beta}\alpha}
 \quad ,
\end{equation}
implies the algebraic relation
\begin{equation}\label{eq1.3}
    p_\mu p^\mu = p_{\alpha \dot{\beta}}\, p^{\dot{\beta}\alpha }\, =
    |M|^2 \quad ,
\end{equation}
where the Lorentz-invariant bilinear\footnote{In
\cite{10azluk,11azluk} the variable $f=\lambda_{\alpha 1}
\lambda^{\alpha}_{\  2} = {\textstyle \frac{1}{\sqrt{2}}} M$ was
used in place of $M$.}
\begin{equation}\label{eq1.4}
    M= {\textstyle \frac{1}{\sqrt{2}}}
    \lambda_{\alpha i} \, \lambda^{\alpha i}\quad
    (\, \bar{M}= {\textstyle \frac{1}{\sqrt{2}}}
    {\bar{\lambda}}_{\dot{\alpha} i} \, {\bar{\lambda}}^{\dot{\alpha}
    i} \,)
\end{equation}
gives the composite complexified mass and breaks the conformal
invariance down to the Poincar\'e one. To obtain a real mass $m$ it
suffices to consider a twistor theory invariant under the phase
transformations
\begin{equation}\label{eq1.5}
    \lambda'_{\alpha i} = e^{i \varphi} \, \lambda_{\alpha i}
    \quad , \quad
    \bar{\lambda}'_{\dot{\alpha} i}=
     e^{- i \varphi} \, {\bar \lambda}_{\dot{\alpha} i}\quad ;
\end{equation}
by a suitable gauge fixing, a real $M \to |M| = m$ is obtained.

\indent
   Twistorial particle models constructed from several twistors
are known \cite{3azluk,Perjes,7azluk,8azluk}; in particular a
two-twistor model was proposed recently to describe free
relativistic particles with mass and spin
\cite{10azluk}-\cite{12azluk}. These considerations confirm that
the introduction of a non-vanishing Pauli-Luba\'nski vector,
describing the relativistic spin fourvector in terms of twistor
coordinates, requires at least two twistors \cite{8azluk}. In this
paper we extend the two-twistor particle dynamics by considering
two $N=2$ supertwistors to describe the degrees of freedom of our
elementary system. The supersymmetric two-twistor geometry will be
arranged in a way that leads algebraically to the condition
\begin{equation}
\label{Xivalue} \Xi=M \; ,
\end{equation}
where $\Xi$ is the complex central charge of the $D$=4, $N$=2
superPoincar\'e algebra. We shall use the phase symmetry
(\ref{eq1.5}) so that the value $M$ of the central charge $\Xi$ in
(\ref{Xivalue}) becomes a real parameter. An interesting outcome
of our approach will be the description of the fermionic
$\kappa$-gauge transformations in a purely twistorial formulation
of the massive superparticle model. Further, it may be shown that
the presence of the mass in the twistorial framework reduces the
conformal supersymmetry realized in terms of the two $N$=2
supertwistors to the $N$=2 superPoincar\'{e} symmetry with a
composite central charge given by (\ref{eq1.4}).

\indent The plan of the paper is the following:
      In Sec. 2 we introduce a massive
two-twistor particle model without supersymmetry using a hybrid
spinor/spacetime formulation. We also present there two other
equivalent spacetime and purely twistorial geometry formulations
and exhibit their relation with the associated six-dimensional
massless particle model. In Sec. 3 we introduce the $N=2$
supersymmetric extension of the bosonic massive model, with
degrees of freedom described by two ($i=1,2$) $N=2$ ($r=1,2$)
 supertwistors \cite{13azluk}
\begin{equation}
\label{eq1.7}
{\cal Z}_{Ri} = (Z_{Ai}, \eta_{ri} )
 =(\lambda_{\alpha i}, {\bar \omega}^{\dot{\alpha}}_{\ \  i}, \eta_{r i} ) \; ,
\end{equation}
where $Z_{Ai}$ and $\eta_{ri}$ are, respectively, Grassmann even
and odd. This supersymmetric model will be formulated first in a
hybrid spinor/superspace geometric framework \cite{5azluk}. We
discuss ({\it cf.} \cite{15azluk,16azluk}) the constraints and, in
particular, the fermionic first class constraints that generate
the $\kappa$-transformations with odd gauge parameters
\cite{17azluk,Sie83}.

\indent Further, present in Sec. 3 a purely supertwistorial
(two-supertwistor) formulation of our model. Its
lagrangian will be the sum of two terms, \\
i) a first one describing the free two-supertwistor lagrangian,
and\\
ii) an additional bilinear term in the fermionic sector which
couples the two supertwistors; this will turn out to be a
supertwistorial Wess-Zumino (WZ) term.

\indent The model is completed by adding the Lagrange multipliers
that describe its algebraic constraints.

\indent   Sec. 4 is devoted to quantizing the fermionic sector of
the two supertwistor model, to uncover the quaternionic structure
behind it and to calculating the fermionic gauge
$\kappa$-transformations. These $\kappa$-symmetries reduce by half
(from four to two) the number of complex Grassmannian
supertwistorial degrees of freedom. We find new features of the
two-supertwistor formulation with respect to single supertwistor
models: the appearance of a WZ term and the presence of the
fermionic $\kappa$-gauge transformations associated with the
non-physical fermionic degrees of freedom in the
multi-supertwistorial formulation.  If we impose the quaternionic
$SU(2)$ Majorana condition in the fermionic sector, the redundant
degrees of freedom of the two supertwistor coordinates described
in the fermionic sector by the $\kappa$-transformations can be
eliminated. Subsequently, we show also in Sec. 4 that it is
possible to introduce $\kappa$-invariant fermionic variables which
describe the fermionic sector of our model in terms of two (rather
than four) complex odd degrees of freedom. It is then seen that
these $\kappa$-invariant fermionic variables can be interpreted as
the dimensionally reduced ($D=6 \rightarrow D=4$) odd components
of a six-dimensional complex supertwistor, satisfying
$SU(2)$-Majorana conditions.

\indent Sec. 5 exploits the above six-dimensional interpretation
of the two-supertwistor model, using a pair of $N=1$, $D=4$
(super)twistors to describe a single $D=6$ (super)twistor. We
perform the $D=6 \to D=4$ dimensional reduction by taking the
six-momentum coordinates $p_4$, $p_5$ as constants. As a result,
we recover the $D$=4 $N$=2 massive superparticle model formulated
by two of the present authors a quarter century ago, the first one
with mass in the fermionic sector introduced by means of a WZ term
\cite{17azluk}. \indent The last section provides an outlook. In
particular, it is argued there that using $N$ $D$=4 supertwistors
we can extend our construction and obtain a $N$-supertwistorial
model with $N$ supersymmetries as well as with $\frac{N(N-1)}{2}$
complex composite mass-like parameters playing the role of central
charges. In particular, the $D=4, N=4$ model would be especially
interesting, since its $\kappa$-invariant formulation could be
given by the coordinates of a single $D=10$ supertwistor with
octonionic structure.

 \section{Massive particle model from $D=6$ and its
 $D=4$ twistorial picture}

           \setcounter{equation}{0}
  It is known that $D=4$ massive free particle models can be obtained
by dimensional reduction from massless particle ones in $D>4$.
Because the twistorial mass (\ref{eq1.4}) is complex, we shall
take $D=6$ and denote $M=p_4 - ip_5$, $\bar{M}=p_4 + ip_5$
($(p_\mu , p_4 , p_5)$ are the six-momentum coordinates) and,
similarly, $z=x^4 +ix^5$, $\bar{z}= x^4 -ix^5$. In the first order
formalism, the Lagrangian of a $D$=6 massless particle can be
written as
\begin{equation}
\label{eq2.1}
{\cal L}^B = p_\mu \dot{x}^\mu + \frac{1}{2}
(M\dot{z} + \bar{M} \dot{\bar{z}}) +
 \frac{e}{2}(p^2 - |M|^2) \; ,\; \mu=0,1,2,3\quad.
\end{equation}
Using eqs. (\ref{eq1.2}) and (\ref{eq1.4}) the `mixed' or
spinor/spacetime formulation of (\ref{eq2.1}) is obtained, in
which
\begin{equation}
\label{eq2.2}
 {\cal L}^B = \bar{\lambda}_{\dot{\beta} i} \,
\dot{x}^{\dot{\beta}\alpha }\,\lambda_{\alpha i} +
\frac{1}{2\sqrt{2}} \left( \lambda_{\alpha i} \, \lambda^{\alpha
i} \dot{z} + \bar{\lambda}_{\dot{\alpha} i} \,
\bar{\lambda}^{\dot{\alpha} i} \dot{\bar{z}} \right)
  \, ,
\end{equation}
where the spacetime vector is expressed as a second rank spinor,
 $x_\mu = {\textstyle \frac{1}{\sqrt{2}}} \,
 (\sigma_{\mu})^{{\dot\beta}{\alpha}} x_{\alpha
\dot{\beta}}$ (see the first footnote) and the $D$=6 zero mass
shell condition, $p^kp_k= p^\mu p_\mu - |M|^2= 0$
($k=0,1,\dots,5$), is omitted because it is an identity in terms
of the spinor variables $\lambda_{\alpha i},
\bar{\lambda}_{\dot{\alpha}i}$. Clearly, the lagrangian
(\ref{eq2.2}) is invariant under the rigid phase transformations
(\ref{eq1.5}) if the $z$'s have a ({\it cf.} (\ref{eq1.5})) double
$U(1)$ charge,
\begin{equation}
\label{eq2.3} z' = e^{-2i \varphi} z \qquad ,\qquad \bar{z}' =
e^{2i \varphi} \bar{z} \quad .
\end{equation}

\indent  To obtain a purely twistorial description of the
lagrangian (\ref{eq2.2}) we now introduce a new Weyl spinor and
its conjugate ($\omega^\alpha_i$, ${\bar \omega}^{\dot \alpha}_i$)
and postulate for the components of the twistors $Z_{Ai}=
(\lambda_{\alpha i}, {\bar \omega}^{\dot \alpha}_i)$ the following
incidence relations:
  \begin{eqnarray}
  \label{eq2.6}
    \omega^{\alpha }_{\ i}
     &=&
     i (\bar{\lambda}_{{\dot \beta} i}\,x^{ \dot{\beta}\alpha} \;
     + {\textstyle \frac{1}{\sqrt{2}}} \lambda^{\alpha i}\; z)\;,
      \cr
\bar{\omega}^{\dot{\alpha }}_{\ i} &=&
     -i( x^{\dot{\alpha}\beta}\,{\lambda}_{\beta i}
     +  {\textstyle \frac{1}{\sqrt{2}}} \ \bar{\lambda}^{\dot{\alpha}i }\; \bar{z}
)\; ,
  \end{eqnarray}
$(x^{{\dot \alpha} \beta})^\dagger = x^{{\dot\beta}\alpha}$
($x^\mu$ real). Eqs. (\ref{eq2.6}) generalize the well known
Penrose formula \cite{1azluk}-\cite{4azluk}. Using them, the
lagrangian (\ref{eq2.2}) can be written as
  \begin{eqnarray}
  \label{eq2.7}
  {\cal L}^B & = &
  - {\textstyle \frac{i}{2}}
  \left(\dot{\omega}^{\alpha } _{\ i } \lambda_{\alpha i} +
  \dot{\bar{\lambda}}_{\dot{\alpha}i} \bar{\omega}^{\dot{\alpha}}_i \;
  \right)
+   {\textstyle \frac{i}{2}}
   \left( \omega^{{\alpha}}_{\ i} \; \dot{\lambda}_{\alpha i}
  + \bar{\lambda}_{\dot{\alpha} i} \; \dot{{\bar\omega}}^{\dot{\alpha} }_{\ i }\right)
  \cr
  & = & {\textstyle \frac{i}{2}}
 \left(\bar{Z}^{A}_{\ i} {\dot{Z}}_{A i}-\dot{\bar Z}^{A }_{\ i} Z_{A i} \right)\quad,
 \end{eqnarray}
 where the scalar product of two twistors is given by
 \begin{equation}\label{eq2.8}
   \bar{Z}^{A }_{\ i}\; Z_{Ak} = \omega^{\alpha }_{\ i} \ \lambda_{\alpha k} \;
  + \; \bar{\lambda}_{\dot{\alpha} i} {\bar \omega}^{\dot{\alpha}} _{\ k}
  \; ,
 \end{equation}
 with $\bar{Z}^A_i =({\omega^\alpha}_i ,\bar{\lambda}_{\dot{\alpha} i})$.

\indent One can check, using eqs. (\ref{eq2.6}) and the relations
 \begin{equation}\label{eq2.9}
    \lambda_{\alpha i} \; \lambda^{\alpha} _{\ j} =
    - {\textstyle \frac{1}{\sqrt{2}}} \;
    \epsilon_{ij} \; M \quad ,
    \qquad
     \bar{\lambda}_{\dot{\alpha}  i} \; \bar{\lambda}^{\dot{\alpha} } _{\ j} =
     - {\textstyle \frac{1}{\sqrt{2}}} \;
    \epsilon_{ij} \; \bar{M} \quad ,
 \end{equation}
 that
\begin{equation}
 \label{eq2.10}
 \bar{Z}^{A}_{\ i}\, Z_{Aj} = \frac{i}{2}
 \left(Mz-\bar{M}\bar{z}\right) \delta_{ij} \; .
\end{equation}
Since $\bar{Z}^{A}_{\ k}\, Z_{Ak} =
i\left(Mz-\bar{M}\bar{z}\right)$, it follows that  the $z,\bar{z}$
coordinates may be removed by the relations (\ref{eq2.10}), which
can be rewritten as (see also \cite{6azluk})
\begin{equation}
 \label{eq2.10b}
\bar{Z}^{A}_{\ i}\, Z_{Aj} -
 \frac{1}{2}\,\delta_{ij}\, \bar{Z}^{A}_{\ k}\, Z_{Ak}\, = 0  \; ,
\end{equation}
an expression that does not fix the conformal norm of the twistors
$Z_{Ak}$ but states that the two norms are equal.

 \indent
 Summarizing, ${\cal L}^B$ in eq. (\ref{eq2.7}) appears as
 the sum of two free twistor  lagrangians, associated with two
 non-null, orthogonal twistors having the same non-vanishing
 length for $M\neq 0$. To specify that the purely twistorial
 lagrangian (\ref{eq2.7}) describes a massive particle one
 has to incorporate relations (\ref{eq2.10b}) and (\ref{eq1.4})
 by means of suitable Lagrange multipliers.

 \section{Two-supertwistor model with $D=4,N=2$  supersymmetry and mass}
   \setcounter{equation}{0}

   {\it a) Supersymmetrization of the model (\ref{eq2.2}) and
$\kappa$-transformations.}

\indent The `hybrid' or spinor/spacetime model (\ref{eq2.2}) is
supersymmetrized
 by replacing the translation-invariant differentials
 $dx^{\alpha \dot{\beta}}$, $dz$ and $d\bar{z}$ by the
corresponding supertranslation-invariant ones on $D=4$, $N=2$
superspace extenden by a central charge and parametrized
 by ($x_{\alpha \dot{\beta}}, z, \bar{z};\ \theta_{\alpha r},
\bar{\theta}_{\dot{\alpha} r}\,$), $r=1,2$. This is achieved by
the replacements\footnote{The
   $\sqrt{2}$ in (\ref{eq3.1a}) is needed because in
   this way $\omega^\mu(\tau)={\textstyle\frac{1}{\sqrt{2}}}
   \sigma^\mu_{\alpha{\dot\beta}}{\omega}^{{\dot\beta}\alpha}(\tau)
   ={\dot x}^\mu - i({\dot{\theta}}^\alpha_r
   \sigma^\mu_{\alpha{\dot\beta}}{\bar\theta}^{\dot
   \beta}_r-\theta^\alpha_r\sigma^\mu_{\alpha{\dot\beta}}{\dot{\bar\theta}}^{\dot
   \beta}_r)$ is the pull-back to to the worldline of the particle
   of the superspace Maurer-Cartan one-form
   $\Pi^\mu= dx^\mu-i(d\theta^\alpha_r
   \sigma^\mu_{\alpha{\dot\beta}}{\bar\theta}^{\dot
   \beta}_r-\theta^\alpha_r\sigma^\mu_{\alpha{\dot\beta}}d{\bar\theta}^{\dot
   \beta}_r)$, which is invariant under the $D=4,N=2$
   superPoincar\'e transformations in eqs. (\ref{eq3.3}).}
\renewcommand{\theequation}{\arabic{section}.1\alph{equation}}
\setcounter{equation}{0}
   \begin{eqnarray}
   \label{eq3.1a}
        \dot{x}^{\dot{\beta}\alpha} \to
        \omega^{{\dot\beta}\alpha} =
        \dot{x}^{\dot{\beta}\alpha} -
        i\sqrt{2}\left(
        \dot{\theta}^\alpha_{\ r} \; {\bar\theta}^{\dot{\beta}}_{\ r}
        - \theta^{\alpha}_{\ r } \; \dot{\bar{\theta}}^{\dot{\beta}}_{\ r}
        \right)\;,\quad\quad
        \\
        \dot{z} \to \omega = \dot{z} + 2 i\;\theta_{\alpha r}
        \dot{\theta}^{\alpha r}
        \quad , \quad
        \dot{\bar{z}} \to \bar{\omega} = \dot{\bar{z}}
         + 2 i \;  {\bar\theta}_{\dot{\alpha} r} \; \dot{\bar \theta}^{\dot{\alpha} r} \quad ,
\label{eq3.1b}
   \end{eqnarray}
where $a_r b^r \equiv a_r \epsilon^{rs} \, b_s$  and $z,\bar{z}$
play the r\^ole of the $D$=4, $N$=2 superspace
 complex central charge coordinates. These replacements in (\ref{eq2.2}) give
 the supersymmetric lagrangian
\renewcommand{\theequation}{\arabic{section}.\arabic{equation}}
\setcounter{equation}{1}
\begin{equation}
\label{eq3.2}
    {\cal L}^{SUSY} =
      \bar{\lambda}_{\dot{\beta} i}
    \; \omega^{\dot{\beta}\alpha}\,\lambda_{\alpha  i}
    + \frac{1}{2\sqrt{2}} \left( \lambda_{\alpha  i}\; \lambda^{\alpha  i} \; \omega +
\bar{\lambda}_{\dot{\alpha}  i}\; \bar{\lambda}^{\dot{\alpha}  i} \;
\bar{\omega} \right) \; .
\end{equation}

\indent  The action obtained from (\ref{eq3.2}) is invariant under
the supertranslations of the $N=2$ superPoincar\'{e} group
extended by a complex central charge,
\begin{eqnarray}
\label{eq3.3} \nonumber
 && x'^{{\dot \beta}\alpha} = x^{{\dot
\beta}\alpha} -i\sqrt{2}(\epsilon^{\alpha} _{\ r}\;
\bar{\theta}^{\dot{\beta}}_{\ r} - \theta^{\alpha}_{\ r} \;
\bar{\epsilon}^{\dot{\beta}}_r )\; ,
\\
\nonumber
 && \theta'_{\alpha r} = \theta_{\alpha r} +
\epsilon_{\alpha r} \quad \quad ,\quad \bar{\theta}'_{\dot{\alpha}
r} = \bar{\theta}_{\dot{\alpha} r}
 + \bar{\epsilon}_{\dot{\alpha} r}\; ,
\\
\nonumber
&& z' = z +2 i \theta_{\alpha r}\; \epsilon^{\alpha r}
\quad , \quad \bar{z}' = \bar{z} + 2 i
\;\bar{\theta}_{\dot{\alpha} r}\; \bar{\epsilon}^{\dot{\alpha}r}
\; ,
\\
&& {\lambda'}_{\alpha i} = {\lambda}_{\alpha i}\qquad \;
\qquad,\qquad {\bar{\lambda}'}_{\dot{\alpha} i} =
{\bar{\lambda}}_{\dot{\alpha} i} \quad ,
\end{eqnarray}
which leave ${\lambda}_{\alpha i}$ invariant. Expression
(\ref{eq3.1a}) is also invariant under the $U(2)$ internal
transformation of the odd superspace coordinates
 $\theta_{\alpha r}, \theta_{\dot{\alpha} r}$; this symmetry is
 broken by the $\omega$'s of eq.(\ref{eq3.1b}) down to the
 $U(2)\cap Sp(2,C)=USp(2)\approx SU(2)$ internal
 symmetry.

 \indent  The lagrangian (\ref{eq3.2}) describes a superparticle in a mixed
spinorial/superspace configuration space ${\cal M}^{(6; 8|8)}$
parametrized by
 \renewcommand{\theequation}{\arabic{section}.\arabic{equation}}
\setcounter{equation}{3}
\begin{equation}\label{eq3.4}
    {\cal M}^{(6; 8|8)} = \{ q^{\cal M} \} =
   ( x^{{\dot \alpha}\beta}, z, \bar{z};
    \lambda^{\alpha i} , {\bar \lambda}^{\dot{\alpha} i}
    | \theta^{\alpha}_r, \bar{\theta}^{\dot{\alpha}}_r)\quad ,
\end{equation}
with 4+2+8 =14 real bosonic and 8 real fermionic coordinates. The
canonical momenta ($\pi_{\alpha r}= {\partial L}/\partial {\dot
\theta}^\alpha_r$, etc.)
\begin{equation}
\label{eq3.5}
    {\cal P}_{\cal M} = \frac{\partial L}{\partial \dot{q}^{\cal M}}
    \equiv  \left( p_{{\dot \alpha}\beta}, q, \bar{q}; \rho_{\alpha
i}, \bar{\rho}_{\dot{\alpha}  i} \; | \pi_{\alpha r},
\bar{\pi}_{{\dot{\alpha}} r} \right)\; ,\;{\cal M}= 1 \ldots 22
\quad,
\end{equation}
define the following set of primary constraints:
\begin{eqnarray}\label{eq3.6}
R_{\alpha \dot{\beta}} & := &
 p_{\alpha \dot{\beta}} - \lambda_{\alpha  i}
 \; \bar{\lambda}_{\dot{\beta} i} = 0 \quad ,
 \cr
 R\; & := &
 {q} - {\textstyle \frac{1}{2\sqrt{2}}}
 \lambda_{\alpha  i}
 \; {\lambda}^{{\alpha} i} =
 q - {\textstyle \frac{M}{2}}=  0 \quad ,
 \cr
 \bar{R}\; & := &
 \bar{q} -  {\textstyle \frac{1}{2\sqrt{2}}} \; \bar{\lambda}_{\alpha  i}
 \; \bar{\lambda}^{{\alpha} i} =
q - {\textstyle \frac{\bar{M}}{2}} = 0 \quad ,
 \cr
 R_{\alpha  i}
  & := & \rho_{\alpha  i} = 0 \quad , \quad
 R_{\dot{\alpha}  i} :=
  \bar{\rho}_{\dot{\alpha}  i} = 0 \quad ,
\end{eqnarray}
and  ($\bar{G}_{{\dot \alpha} r}$=$-(G_{\alpha r})^+$,
$\bar{\pi}_{{\dot \alpha} r}$=$-(\pi_{\alpha r})^+)$
\renewcommand{\theequation}{\arabic{section}.7\alph{equation}}
\setcounter{equation}{0}
\begin{eqnarray}
\label{eq3.7a} G_{\alpha r} & := & \pi_{\alpha r} + i{\textstyle
\sqrt{2}} p_{\alpha \dot{\beta}}\; \bar{\theta}^{\dot{\beta}}_{\
r} - i M \; \epsilon_{rs} \; \theta_{\alpha s} = 0 \quad ,
\\\label{eq3.7b}
\bar{G}_{\dot{\alpha} r} & := & {\bar \pi}_{\dot{\alpha} r} +
i{\textstyle \sqrt{2}} \theta^{\beta}_{\ r} p_{\beta \dot{\alpha}} -
 i\bar{M} \; \epsilon_{rs} \; \bar{\theta}_{\dot{\alpha} s} = 0 \quad .
\end{eqnarray}
\renewcommand{\theequation}{\arabic{section}.\arabic{equation}}
\setcounter{equation}{7}

\indent   Let us restrict ourselves to the set (\ref{eq3.7a}),
(\ref{eq3.7b}) of fermionic constraints, which determine the
elements of the Poisson brackets (PB) matrix
\begin{equation}\label{eq3.8}
    {\cal C}_{AB} =
    \left(
    \begin{array}{ll}
\{ G_{\alpha r}, G_{\beta s} \}, & \{ G_{\alpha r},
\bar{G}_{\dot{\beta} s}\} \cr \{\bar{G}_{\dot{\alpha} r}, G_{\beta
s}\},& \{ \bar{G}_{\dot{\alpha} r}, \bar{G}_{\dot{\beta} s}\}
    \end{array}
    \right) \quad .
\end{equation}
Using the canonical PB
\begin{equation}\label{eq3.9}
    \{ \theta_{\alpha r} \, ,\pi_{\beta s}  \}
    =  \epsilon_{\alpha \beta} \; \delta_{rs}
\quad , \quad
    \{ \bar{\theta}_{\dot{\alpha} r}\,,\,\bar{\pi}_{\dot{\beta} s}\}
    =  \epsilon_{\dot{\epsilon} \dot{\beta}} \delta_{rs} \quad ,
\end{equation}
it follows that the four 4$\times$4 blocks of the ${\cal C}_{AB}$
matrix are given by ($p _{\dot{\beta} \alpha} = (p_{\alpha
\dot{\beta}})^T$)
\begin{equation}\label{eq3.10}
    {\cal C}_{AB} =2i \left(
    \begin{array}{cc}
    - \epsilon_{\alpha \beta}\; \epsilon_{r s} M &
    \sqrt{2}\delta_{rs} \; p_{\alpha \dot{\beta}}
    \cr
    \sqrt{2}\delta_{rs} \; p_{\beta \dot{\alpha} }
    &
     - \epsilon_{\dot{\alpha} \dot{\beta}}\; \epsilon_{r s} \bar{M}
     \end{array}
    \right) \quad .
\end{equation}
Using the formula for the determinant of a $2\times 2$
 blocks matrix,
\begin{eqnarray}
 \label{eq3.11a}
\det {\cal C} & = &
 \det A \cdot \det(D - CA^{-1} B) \quad , \quad
 {\cal C} = \left(
\begin{array}{cc} A & B \cr C & D
\end{array} \right)\; ,
\end{eqnarray}
one finds that
\begin{eqnarray}
\label{eq3.11b}
    \det {\cal C} & = &  2^8(p^2 - |M|^2 )^4 \quad.
\end{eqnarray}
\\ \noindent
The first constraint in (\ref{eq3.6}) reproduces eq.
(\ref{eq1.2}), implying
\begin{equation}
\label{eq3.12}
    p^2 - |M|^2 = 0
\end{equation}
(eq. (\ref{eq1.3})), and thus we conclude from (\ref{eq3.11b})
that the 8$\times$8 PB constraints matrix (\ref{eq3.10}) is of
rank four.

\indent  We may now derive four first class fermionic constraints
by multiplying respectively (\ref{eq3.7a}) by $p^{\alpha
\dot{\gamma}}$ and (\ref{eq3.7b}) by $\epsilon_{rs}\; M$. This
gives
\begin{equation}
\label{eq3.13}
   {\bar C}_{{\dot \beta} r}
= \pi^{\alpha}_r \, p_{\alpha \dot{\beta}} \;  +
\frac{M}{\sqrt{2}}\epsilon_{rs } \,
     \bar{\pi}_{\dot{\beta} s} - \frac{i}{\sqrt{2}}
     \bar{\theta}_{\dot{\beta} r} (p^2-|M|^2)= 0 \, .
\end{equation}
Equation (\ref{eq3.13}) determines in principle four complex
constraints, but their complex-conjugate ones are equivalent to
them. Indeed, if we multiply the constraints ($C_{\beta
r}^+=-({\bar C}_{\dot{\beta} r}) $)
\begin{equation}\label{eq3.14}
{C}_{\beta r}=
    p_{\beta \dot{\alpha}} \; {\bar\pi}^{\dot{\alpha}}_r
    + \frac{\bar{M}}{\sqrt{2}}\epsilon_{rs} \,
     {\pi}_{\beta s}-\frac{i}{\sqrt{2}}\theta_{\beta r} (p^2- |M|^2) = 0 \; ,
\end{equation}
by $p^{{\dot\gamma}\beta}$ we get back the constraints
(\ref{eq3.13}), plus terms that contain the factor $(p^2-|M|^2)$.
Therefore our model has effectively only four real first class
constraints, which in the 8-dimensional real Grassmann odd sector
of the configuration space (\ref{eq3.4}) generate four real odd
gauge transformations. These are the $\kappa$-symmetries of the
model (\ref{eq3.2}) that eliminate the unphysical fermionic gauge
degrees of freedom {\it i.e.}, half of the odd $N$=2 superspace
coordinates. The explicit expression of these
$\kappa$-transformations, parametrized by a pair of anticommuting
Weyl spinors $\kappa_{\alpha r}$ and their complex conjugates
${\bar{\kappa}}_{\dot{\alpha}r}$, is given by the graded Poisson
brackets
\begin{eqnarray}
\label{eq3.15x}
\delta_\kappa \; \theta^{\alpha}_r & := &
  \{ \kappa^{\beta}_s C_{\beta s}  , \theta^{\alpha}_r \} =
  \kappa^\beta_s \{ C_{\beta s}  , \theta^{\alpha}_r \} \; =
  - \epsilon_{rs} \frac{\bar M}{\sqrt{2}} \, \kappa^{\alpha}_s \;
  ,\cr\cr
\delta_{\bar{\kappa}} \theta_{ \alpha r}  & :=&
  \{  \bar{\kappa}^{\dot{\beta}}_s \bar{C}_{\dot{\beta}  s} , { \theta}_{ \alpha r} \}
  = \bar{\kappa}^{\dot{\beta}}_s \{ \bar{C}_{\dot{\beta}  s},
  {\theta}_{ \alpha r} \} = -p_{\alpha \dot{\beta}}
  \bar{\kappa}^{\dot{\beta}}_r\;  \quad, \cr\cr
\delta_\kappa {\bar \theta}^{\dot \alpha}_r & := &
  \{  \kappa^{\beta}_s C_{\beta  s} , {\bar \theta}^{\dot \alpha}_r \} \;
  = \; \kappa^{\beta}_s \{ C_{\beta s} , {\bar \theta}^{\dot \alpha}_r
  \} \; = \;\,  p^{\dot{\alpha}\beta }\; \kappa_{\beta r}\; , \cr\cr
\delta_{\bar{\kappa}} \; \bar{\theta}^{\dot{\alpha}}_r  &:=&
 \{ \bar{\kappa}^{\dot{\beta}}_s \bar{C}_{\dot{\beta} s}  ,
 \bar{\theta}^{\dot{\alpha}}_{r } \} =
 \; \bar{\kappa}^{\dot{\beta}}_s  \{ \bar{C}_{\dot{\beta} s}  ,
 \bar{\theta}^{\dot{\alpha}}_{r }\} \; =
  - \epsilon_{rs} \frac{ M}{\sqrt{2}} \, \bar{\kappa}^{\dot{\alpha}}_s
\quad.
\end{eqnarray}
If $\delta_\kappa$ now denotes the variation under both $\kappa$
and $\bar{\kappa}$, the variation of the four-dimensional spinor
$(\theta_{\alpha r}, \bar{\theta}^{\dot{\alpha}}_r)$ is written as
\begin{equation}
\label{3.15b}
   \delta_\kappa \left( \begin{array}{c} \theta_{\alpha r} \cr
   \bar{\theta}^{\dot{\alpha}}_r  \end{array} \right) =
   \left(  \begin{array}{cc}  -\epsilon_{rs} \delta^\beta_\alpha
   \frac{\bar{M}}{\sqrt{2}} & - \delta_{rs} p_{\alpha\dot{\beta}} \cr
   \delta_{rs} p^{\dot{\alpha}\beta} & -\epsilon_{rs}
   \delta^{\dot{\alpha}}_{\dot{\beta}} \frac{M}{\sqrt{2}} \end{array} \right)
   \left( \begin{array}{c} \kappa_{\beta s} \cr
   \bar{\kappa}^{\dot{\beta}}_s  \end{array} \right)
    \ .
\end{equation}
The above matrix can be rewritten as the product
\begin{equation}
\label{3.15bc}
\left(  \begin{array}{cc}  -\epsilon_{rt} \delta^\gamma_\alpha
   \frac{\bar{M}}{\sqrt{2}} & 0 \cr
   0 & -\epsilon_{rt}
   \delta^{\dot{\alpha}}_{\dot{\gamma}} \frac{M}{\sqrt{2}} \end{array} \right)
   \left(  \begin{array}{cc}  \delta_{ts} \delta^\beta_\gamma
    & - \epsilon_{ts} \frac{\sqrt{2}}{\bar{M}} p_{\gamma\dot{\beta}} \cr
   \epsilon_{ts} \frac{\sqrt{2}}{M} p^{\dot{\gamma}\beta} & \delta_{ts}
   \delta^{\dot{\gamma}}_{\dot{\beta}}  \end{array} \right)\ .
\end{equation}
The first matrix just produces a scaling of the
$\kappa$-transformations, and the second matrix is the sum of the
four-dimensional unit matrix plus one with only non-zero $2\times
2$ antidiagonal boxes that squares to one {\it i.e.}, it is a
projection operator. Thus, $\delta_\kappa \theta_{r}$ in eq.
(\ref{3.15b}) has the standard projector structure effectively
halving the parameters of the $\kappa$-transformations.

\indent The behaviour of the remaining configuration space
variables (\ref{eq3.4}) under $\delta_\kappa$ is given by
\begin{eqnarray}
\label{eq3.16x}
\delta_\kappa  x^{\dot{\beta}\alpha} & = &
  i\sqrt{2}( \delta_\kappa \theta^{\alpha}_{\ r} \;
\bar{\theta}^{\dot \beta}_{\ r}
 - \theta^{\alpha}_r \; \delta_{\kappa}{\bar \theta}^{\dot \beta}_{\ r}
 )\ ,
\cr \delta_\kappa {z} & = & -2i {\theta}_{\alpha r} \;
\delta_\kappa \; {\theta}^{\alpha r} \ ,\cr
 \delta_\kappa \bar{z}& = & -2i \bar{\theta}_{\dot{\alpha} r} \; \delta_\kappa \;
\bar{\theta}^{\dot{\alpha} r}\ , \cr \delta_\kappa \lambda_{\alpha
i} & = &
 \delta_\kappa \; \bar{\lambda}_{\dot{\alpha} i} = 0 \quad .
\end{eqnarray}
These relations differ from eqs. (\ref{eq3.3}) by the replacement
$\epsilon \rightarrow -\delta_\kappa \theta$; the relative minus
sign characterizes $\kappa$-symmetry as a `right' (local)
supersymmetry (see {\it e.g.} \cite{AIM04}). The
$\kappa$-transformations (\ref{eq3.15x}), (\ref{eq3.16x}), may be
used to check explicitly the $\kappa$-invariance of the action
based on (\ref{eq3.2}).
\\[12pt]
{\it b)  From the hybrid (spinor/spacetime) formulation to the
purely super twistorial one.}

\indent   To introduce a purely supertwistorial formulation of the
model (\ref{eq3.2}), eqs. (\ref{eq2.6}) are further extended in
the two supertwistors case by\footnote{See \cite{Uvarov07} for
similar relations in the framework of $D$=6 Lorentz harmonics.}
\begin{eqnarray}
\label{eq3.15}
\omega^{\alpha  }_{i} &=& i
\bar{\lambda}_{\dot{\beta} i}( x^{\dot{\beta}\alpha} - i
\sqrt{2}\theta^{\alpha}_{\ r} \; \bar{\theta}^{\dot{\beta}}_{\ r})
 + \frac{i}{\sqrt{2}}\ \epsilon^{ij} ( \lambda^{\alpha }_{\ j}\; z
 + 2i \lambda_{\beta j} \; \theta^{\alpha r}
 \theta^{\beta}_r )\; ,
 \cr
{\bar \omega}^{\dot{\alpha}  }_{i} &=& -i ( x^{\dot{\alpha}\beta}
+ i\sqrt{2} \theta^{\beta}_{\ r} \; \bar{\theta}^{\dot{\alpha}}_{\
r}) {\lambda}_{\beta i} - \frac{i}{\sqrt{2}} \ \epsilon^{ij} (
\bar{\lambda}^{\dot{\alpha} }_{\ j}\; \bar{z}
 - 2i \bar{\lambda}_{\dot{\beta} j} \; \bar{\theta}^{\dot{\alpha}}_{\ r}
 \bar{\theta}^{\dot{\beta} r}) \, ,
\end{eqnarray}
which are the generalized incidence relations for the bosonic
components of the $Z_{Ai}$ part of ${\cal Z}_{Ri}$ (eq.
(\ref{eq1.7})). These relations, which involve the fermionic
superspace coordinates besides the real spacetime
$x^{{\dot{\alpha}}\beta}$ and complex $z$ variables ({\it cf.} eq.
(\ref{eq2.6})), have to be complemented by those affecting the odd
composite variables $\eta_{ri}$, ${\bar \eta}_{ri}$ that make up
\cite{13azluk} the coordinates triple of the two $D$=4 $N=2$
supertwistors,
\begin{equation}
\label{twost}
{\bar{\cal Z}}^R_{ri}= (\omega^\alpha_i \,,\,
{\bar{\lambda}}_{\dot{\alpha} i}\,, \bar{\eta}_{ri}) \quad, \quad
{\cal Z}_{R\,ri}= (\lambda_{\alpha i}\,,\, {\bar
\omega}^{\dot\alpha}_i \,,\, \eta_{ri}) \quad, \quad i=1,2 \quad ,
\end{equation}
where $r=1,2$ is the $N$=2 supertwistor index. Eqs. (\ref{eq3.15})
are accordingly supplemented by \cite{13azluk}
\begin{equation}
\label{eq3.16}
    \eta_{r i} = \sqrt{2}\theta^{\alpha}_{\ r} \; \lambda_{\alpha  i}
    \qquad, \qquad
    \bar{\eta}_{r i} =\sqrt{2} \bar{\theta}^{\dot{\alpha}}_{\ r}
    \; \bar{\lambda}_{\dot{\alpha}  i} \quad .
\end{equation}
Using (\ref{eq1.4}), the above expressions can be inverted with
the result
\begin{equation}
\label{eq3.17}
    \theta^{\alpha}_{\ r}
    = \frac{\lambda^{\alpha j} \; \eta_{r j}}{M}
    \quad , \quad
    \bar{\theta}^{\dot{\alpha}}_{\ r}
    = \frac{\bar{\lambda}^{\dot{\alpha} j} \; \bar{\eta}_{r
    j}}{\bar{M}} \quad .
\end{equation}
It follows from the definition (\ref{eq3.16}) and from eqs.
(\ref{eq3.3}) that under supersymmetry the odd supertwistor
variables transform as
\begin{equation}
\label{etasusy}
 \delta_\epsilon \eta_{ri}=
\sqrt{2}\epsilon^\alpha_r\lambda_{\alpha i}\quad,\quad
 \delta_{\bar\epsilon} {\bar\eta}_{ri}=
 \sqrt{2}{\bar\epsilon}^{\dot\alpha}_r{\bar\lambda}_{\dot{\alpha}i}\;.
\end{equation}
Finally we note that, in terms of the fermionic composite
coordinates, the $\omega,{\bar \omega}$ components (\ref{eq3.15})
of the two $N=2$ supertwistors can be rewritten as
\begin{eqnarray}
\omega^{\alpha  }_{\ i} &=&\;
 i (\bar{\lambda}_{\dot{\beta} i} \,x^{\dot{\beta}\alpha } \; +
 \frac{1}{\sqrt{2}} \; \lambda^{\alpha i}\,
z ) + ( {\theta}^{{\alpha}}_{\ r} \; \bar{\eta}_{ri}
 -\theta^{\alpha r} \eta^i_r) \quad,
 \cr
 {{\bar \omega}}^{\dot{\alpha}  }_{\ i} &=& -i(x^{\dot{\alpha}\beta} \;
 \lambda_{{\beta} i} \; +  \frac{1}{\sqrt{2}} \;
 \bar{\lambda}^{\dot{\alpha}i}\,\bar{z} )
 -({\bar{\theta}}^{\dot{\alpha}}_{\ r} \; {\eta}_{ri}
 - {\bar{\theta}}^{\dot{\alpha} r} \;  \bar{\eta}^i_r)\quad,
\end{eqnarray}
which are the supersymmetric generalizations of eqs.
(\ref{eq2.6}).

\indent   The supersymmetric extension (\ref{eq3.2}) of the
bosonic model (\ref{eq2.2}) can be written in the form
\begin{eqnarray}
\label{eq3.18}
{\cal L}^{SUSY} & = & {\cal L}^{B}  - i \left( \lambda_{\alpha i} \;
\dot{\theta}^{\alpha}_{\ r} \; \bar{\eta}_{r i} +
\bar{\lambda}_{\dot{\alpha} i} \; \dot{\bar{\theta}}^{{\dot
\alpha}}_{\ r}\; \eta_{r i} \right) \cr && + {i} \left(
M\theta_{\beta r}\; \dot{\theta}^{\beta r} + \bar{M}\;
\bar{\theta}_{\dot{\beta} r}\; \dot{\bar{\theta}}^{\beta r}
\right)\; .
\end{eqnarray}
A calculation now shows (modulo a total time derivative) that,
after introducing (\ref{eq3.17}), the purely supertwistorial
lagrangian for our model reads
\begin{eqnarray}
\label{eq3.19}
{\cal L}^{SUSY} & =&  {\cal L}^{SUSY}_1+ {\cal
L}^{SUSY}_2 \quad,\cr
  {\cal L}^{SUSY}_1& \equiv &{\textstyle
\frac{i}{2}} (\bar{Z}^{A }_{\ i} \;\dot{Z}_{A i} -
\dot{\bar{Z}}^{A}_{\ i}\; Z_{A i} ) \; , \cr
 {\cal L}^{SUSY}_2& \equiv &\frac{i}{\sqrt{2}}\left( \eta_{ri} \;\dot{\bar{\eta}}_{ri} -
\dot{\eta}_{ri} \; \bar{\eta}_{ri} \right) - \frac{i}{\sqrt{2}}
\left(\dot{\eta}_{ri}\; \eta^{r i} + \dot{\bar{\eta}}_{ri}\;
\bar{\eta}^{r i} \right) \; ,\;
\end{eqnarray}
where the scalar product of the
${\bar{Z}}^A_i=(\omega^\alpha_i\,,\,{\bar{\lambda}}_{\dot{\alpha}
i})$  and  $Z_{A i}=(\lambda_{\alpha
i}\,,\,{\bar{\omega}}^{\dot{\alpha}}_i)$ twistors, in which
$\omega^\alpha_i\,,\, \bar{\omega}^{\dot{\alpha}}_i$ are those in
\eqref{eq3.15}, is given by eq. (\ref{eq2.8}). Using eq.
(\ref{eq3.17}) it is seen that the ${\cal L}^{SUSY}_1$ part of
${\cal L}^{SUSY}$ depends only on
$\eta\,,\bar{\eta},\lambda\,,\bar{\lambda}$ and on the time
derivatives of the $\lambda$'s; all the  dependence of ${\cal
L}^{SUSY}$ on the derivatives of the $\eta, \bar{\eta}$ variables
is contained in ${\cal L}^{SUSY}_2$ above.

\indent The $SU(2,2|2)$-invariant product of $N$=2 supertwistors
(\ref{twost}) is defined by\footnote{The different components of
the (super)twistors are not dimensionally homogeneous. In natural
units, $[\lambda]=L^{-\frac{1}{2}}$,
$[{\bar{\omega}}]=L^{\frac{1}{2}}$, $[\eta]=L^0$; the scalar
products of (super)twistors (eqs. (\ref{eq2.8}), (\ref{eq3.20}))
are, of course, dimensionless.
 Note also that (\ref{eq1.7}) implies
that the components $\lambda$, $\bar{\omega}$ and $\eta$ of the
(super)twistors transform in the same manner under a symmetry group
acting on the index $i$ that labels the two (super)twistors.}
\begin{equation}
\label{eq3.20}
\bar{{\cal Z}}^{R }_{ \ i} {\cal Z}_{R j} \; =
\bar{Z}^{A }_{\ i}\; Z_{A j} \; +\sqrt{2} \; \bar{\eta}_{r
i}\;{\eta}_{r j} \quad .
\end{equation}
The bosonic subgroup of $SU(2,2|2)$ is $SU(2,2)\times U(2) \approx
\widetilde{SO}(2,4)\times U(2)$, of which $SU(2,2)$ acts on the
$A$ indices and $U(2)$ on the index $r$; each factor preserves the
two terms in (\ref{eq3.20}) independently. Using (\ref{eq3.20}),
the lagrangian (\ref{eq3.19}) takes the form
\begin{equation}
\label{superact}
{\cal L}^{SUSY} = \frac{i}{2} \left(
 \bar{{\cal Z}}^{R}_{\ i}
\; \dot{{\cal Z}}_{R i} - \dot{\bar{{\cal Z}}}^R_i\; {\cal Z}_{R
i} \right) - \frac{i}{\sqrt{2}} \left( {\dot \eta}_{ri} \; \eta^{r
i} + {\dot {\bar{\eta}}}_{ri} \; {\bar{\eta}}^{ri} \right)\; .
\end{equation}
The first term in (\ref{superact}) is the free lagrangian for two
supertwistors, which are coupled only through the second term.
This last one is the pull-back to the worldline of the
supertwistor space one-form $d\eta\ \eta+d\bar{\eta}\ \bar{\eta}$,
a potential one-form of the closed, supersymmetry-invariant
two-form $d\eta \ d\eta + d{\bar \eta}\ d{\bar \eta}$ and, thus,
${\cal L}^{SUSY}_2$ is the supertwistorial WZ part of ${\cal
L}^{SUSY}$ (for the geometry of WZ rerms, see \cite{WZ}). In fact,
using $\delta_\epsilon \eta$, $\delta_{\bar{\epsilon}} \bar{\eta}$
in (\ref{etasusy}), it is seen that $d\eta \eta+d\bar{\eta}
\bar{\eta}$ is invariant modulo an exact term.

\indent We calculate now, using (\ref{eq3.15}), (\ref{eq3.16})
 and (\ref{eq2.8}) the value of the scalar products (\ref{eq3.20}),
and obtain
\begin{equation}
\label{eq3.21}
    { \bar {\cal Z}}^R_i\; {\cal Z}_{R j}
= \frac{i}{2} \left( Mz - \bar{M}\bar{z} \right) \delta_{ij} -
\frac{1}{\sqrt{2}} \left( \eta_{r i} \; \eta^{r j} + \bar{\eta}_{r
i} \; \bar{\eta}^{r j} \right)\; ,
\end{equation}
({\it cf.} (\ref{eq2.10})). Since $\eta_{r i}\; \eta^{ri} \equiv
0$ due to the $\eta$'s odd Grassmann parity, one obtains
\begin{equation}
\label{eq3.24a}
\; \bar{{\cal Z}}^{R}_k\;{\cal Z}_{R k} = i (Mz -
\bar{M} \bar{z}) \;
\end{equation}
as in the non-supersymmetric case. Proceeding as in Sec. 2 and
using (\ref{eq3.24a}), the constraints (\ref{eq3.21}) may be
rewritten just in terms of supertwistorial variables as
\begin{equation}
 \label{eq3.21b}
    { \bar {\cal Z}}^R_i\; {\cal Z}_{R j}
    - \frac{1}{2} \, \delta_{ij}\, \bar{{\cal Z}}^{R}_k\;{\cal Z}_{R k}
    +\frac{1}{\sqrt{2}}
\left( \eta_{r i} \; \eta^{r j} + \bar{\eta}_{r i} \;
\bar{\eta}^{r j} \right) =0 \; ,
\end{equation}
which extend those in (\ref{eq2.10b}) to the supertwistorial case.
The constraints (\ref{eq3.21b}) and the relations (\ref{eq1.4})
that characterize the model (\ref{superact}) can be incorporated
to it by means of suitable lagrange multipliers.

\indent We now turn to the fermionic gauge symmetries of our
model.

\section{$\kappa$-symmetry and $\kappa$-invariant formulation
 of the fermionic sector}
\renewcommand{\theequation}{\arabic{section}.\arabic{equation}}
\setcounter{equation}{0}

Let us consider now the $L_2$ part of the action (\ref{eq3.19})
involving the derivatives of the fermionic variables. Introducing
new complex Grassmann variables
($\eta^{ri}=\epsilon^{rs}\epsilon^{ij}\eta_{sj}$ etc.)
\begin{equation}
\label{neq4.1} \xi_{ri}  \equiv \eta_{ri} + \bar{\eta}^{ri} \qquad
(\,\bar{\xi}_{ri}  \equiv \bar{\eta}_{ri} + \eta^{ri}\,) \quad,
\end{equation}
and using the Grassmann nature of $\eta_{ri}$, $\bar{\eta}_{ri}$
one gets, up to a total derivative,
\begin{equation}
 \label{neq4.2}
  {\cal L}^{SUSY}_2 =  \frac{i}{\sqrt{2}}
  {\xi}^{ri}\dot{\xi}_{ri} \equiv \frac{i}{\sqrt{2}} \epsilon^{rs}
  \epsilon^{ij} {\xi}_{sj}\dot{\xi}_{ri} \quad .
\end{equation}
If we observe that the variables (\ref{neq4.1}) satisfy the
$SU(2)$-reality condition representing quaternionic structure
\begin{equation}
 \label{neq4.3}
  \xi_{ri}=  \bar{\xi}^{ri} = \epsilon^{rs}
  \epsilon^{ij} \bar{\xi}_{sj} \quad ,
\end{equation}
we see that the action (\ref{neq4.2}) can be written in two
different ways,
\begin{equation}
 \label{neq4.4}
  {\cal L}^{SUSY}_2 =  \frac{i}{\sqrt{2}}
  {\bar{\xi}}_{ri}\dot{\xi}_{ri}=\sqrt{2}\bar{\xi}_i\dot{\xi}_{i}
  \quad .
\end{equation}
where we have chosen $\xi_i\equiv \xi_{1i}$ and used
(\ref{neq4.3}).

\indent Thus, the fermionic action (\ref{neq4.4}) effectively
depends on two complex Grassmann variables only. The
$\kappa$-transformations with Grassmann parameters $\rho_{ri}$
that satisfy the condition ({\it cf.} \eqref{neq4.3})
  such that $\bar{\rho}^{ri}=-\rho_{ri}$
\begin{equation}
\label{neq4.5}
  \delta \eta_{ri} = \rho_{ri} \ , \quad \delta \bar{\eta}^{ri} =
  -\rho_{ri}\quad ,
\end{equation}
leave invariant the variables $\xi_{ri}$ in (\ref{neq4.1}) as well
as the action (\ref{eq3.19}). Using in \eqref{3.15b} the spinor
bilinears (\ref{eq1.2}) for $p_{\alpha\dot{\beta}}$, and comparing
eq. \eqref{3.15b} with (\ref{neq4.5}) one obtains, using eqs.
\eqref{eq1.2}, \eqref{eq2.9} and \eqref{eq3.16},
\begin{equation}
\label{neq4.6}
   \rho_{ri} =-\bar{M}\epsilon_{rs} \lambda^\alpha_i \kappa_{\alpha
   s}+\epsilon_{is} M {\bar{\lambda}}_{\dot{\beta} s}
   \bar{\kappa}^{\dot{\beta}}_r = -{\bar{\rho}}^{ri}
  \quad .
\end{equation}

\indent It is easy to deduce the reality condition (\ref{neq4.3})
if we assume  that the fermionic Grassmann sector of the $D=4$
supertwistor degrees of freedom is described by a single
quaternionic $D=6$ supertwistor coordinate: its Grassmann sector
is given by the odd quaternionic variable $\xi=
\xi_{(0)}+\xi_{(r)} e^r$, where $\xi_{(0)}$, $\xi_{(r)}$
($r=1,2,3$) are four real Grassmann variables
 and $e^r e^s = -\delta^{rs} + \epsilon^{rst}e^t$. In the
matrix representation obtained by replacing the quaternionic units
by the Pauli matrices,
\begin{equation}
\label{neq4.7}
   1\to \sigma^0\quad,\quad e^r \leftrightarrow -i\sigma^r \quad ,
\end{equation}
the quaternionic variable $\xi$  becomes the $2\times 2$ matrix
($\mu=0,r$)
\begin{equation}
   \xi = \sigma^0 \xi_{(0)} -i\sigma^i\xi_{(i)}
   = \left(\begin{array}{cc} \xi_{(0)}
   -i\xi_{(3)} & -i\xi_{(1)} -\xi_{(2)} \\ -i\xi_{(1)} +\xi_{(2)} &
   \xi_{(0)}+i\xi_{(3)}\end{array}\right)\quad .
\label{neq4.8}
\end{equation}
It is trivial to check that $\xi_{ri}$ as given by the elements of
the matrix (\ref{neq4.8}) satisfies the subsidiary condition
(\ref{neq4.3}); clearly, the hermitian matrix $\xi^\dagger$
describes the conjugate quaternion $\xi= \xi_{(0)}-\xi_{(r)} e^r$.
We see therefore that our supertwistor model, described by the
lagrangian (\ref{eq3.19}), reflects the quaternionic structure
inherent to the $D=6$ geometry. The $\kappa$-transformations in
our formulation with two independent complex $D=4$ supertwistors
represent the redundant degrees of freedom which disappear if we
pass to the $N=1$, $D=6$ quaternionic supertwistor coordinates.

\indent The complex Grassmann coordinates $\xi_r$
($=\xi_{1r}\,,\,r=1,2$) satisfy, when the canonical quantization
of the action (\ref{neq4.4}) is performed, the relations ($\hbar
=1$)
\begin{equation}
   \{\xi_r, \bar{\xi}_s \} = \delta_{rs} \ ,\quad
   \{\xi_r, \xi_s \} = \{\bar{\xi}_r, \bar{\xi}_s \} =0 \quad .
\label{neq4.9}
\end{equation}
If we supplement the above anticommutators with the canonical
twistorial equal-time commutators,
\begin{equation}
\label{neq4.10}
    [\lambda_{\alpha i} , \omega^{\beta}_j ]
= i \delta_{\alpha}^{\ \beta} \; \delta_{ij}\quad , \qquad
[\bar{\lambda}_{\dot{\alpha} i} , {\bar{\omega}}^{\dot{\beta}}_j ]
=  i \delta_{\dot{\alpha}}^{\ \dot{\beta}} \; \delta_{ij}\quad ,
\end{equation}
which follow from the symplectic twistorial two-form, we can
postulate the following formulae for the four complex supercharges
describing the algebraic basis of $N=2$, $D=4$ supersymmetry
algebra
\begin{equation}
\label{neq4.11}
      \begin{array}{ll} Q^{(1)}_{\alpha}= \lambda_{\alpha}^{\ i}
      \xi_i\ , & \bar{Q}^{(1)}_{\dot{\alpha}}= \bar{\lambda}_{\dot{\alpha}}^{\ i}
      \bar{\xi}_i \ ,\\ Q^{(2)}_{\alpha}= \lambda_{\alpha i}
      \bar{\xi}_i\ , & \bar{Q}^{(2)}_{\dot{\alpha}}= \bar{\lambda}_{\dot{\alpha
      i}}
      \xi_i\ \ .\end{array}
\end{equation}
Indeed, using the canonical commutation relations (\ref{neq4.9}),
(\ref{neq4.10}) one obtains
\begin{eqnarray} \label{neq5.12} &&\left\{
Q^{(r)}_{\alpha} , \bar{Q}^{(s)}_{\dot{\beta}} \right\}=
\delta_{rs} \; \lambda_{\alpha i} \; \bar{\lambda}_{\dot{\beta} i}
= \delta_{rs} \; P_{\alpha \dot{\beta}} \ ,\cr &&\left\{
Q^{(r)}_{\alpha} , {Q}^{(s)}_{{\beta}} \right\} = \epsilon_{rs} \;
\lambda_{\alpha i} \; {\lambda}_{\beta} ^{\  i} =-
\frac{1}{\sqrt{2}} \;\epsilon_{rs} \; \epsilon_{\alpha {\beta}} M\
, \cr &&\left\{ \bar{Q}^{(r)}_{\dot{\alpha}} ,
\bar{Q}^{(s)}_{\dot{\beta}} \right\} = \epsilon_{rs} \;
\bar{\lambda}_{\dot{\alpha}  i} \; \bar{\lambda}_{\dot{\beta}} ^{\
i} = -\frac{1}{\sqrt{2}} \;\epsilon_{rs} \; \epsilon_{\dot{\alpha}
\dot{\beta}} \; \bar{M}\ ,
\end{eqnarray}
which reproduces the fermionic sector of the $N=2$, $D=4$
supersymmetry algebra with a composite central charge $M$.

\section{The massive $D=4, N=2$  superparticle model with
WZ term from the $D=6$ supertwistorial framework}
\setcounter{equation}{0}

   Our lagrangian (\ref{eq3.2}) may be written in terms of
 $D=6$ four-component Weyl spinors. We introduce
\begin{equation}
\label{eq6.1}
    \Lambda^{ A}_1 =
\left(
\begin{array}{c}
\lambda^{\alpha}_1 \cr  \bar{\lambda}_{\dot{\alpha}  2}
\end{array}
\right) \qquad , \qquad {\Lambda}^{ A}_2 = \left(
\begin{array}{c}
-\lambda^{\alpha}_2 \cr  \bar{\lambda}_{\dot{\alpha}  1}
\end{array}
\right)\ .
\end{equation}
The  complex spinors (\ref{eq6.1}) satisfy the $D$=6 symplectic
Majorana reality condition
\begin{equation}
\label{eq6.2}
\Lambda_{r}^{A} =
 \epsilon^{rs} C^{A\dot{A}} \Lambda_s^{\dot{A}}\; ,
\end{equation}
where $\Lambda^{\dot{A}} \equiv (\Lambda^A)^*$ and
\begin{equation}\label{eq6.3}
    C = \left(
\begin{array}{cc}
0 & \epsilon^{\alpha\beta} \cr -\epsilon_{\dot{\alpha}\dot{
\beta}} & 0
\end{array}
\right) \ ,\qquad C^T = - C \ ,\qquad C^2 = - 1\; ,
\end{equation}
is the charge conjugation matrix.

\indent The $D=6$ generalization of Pauli matrices can be obtained
by replacing in the expression of the $D=4$ Pauli matrices the
imaginary unit by the three quaternionic imaginary units $e^i$
($i=1,2,3$) as follows
\begin{equation}
 \label{neq6.4a}
   \sigma^k = \left( \,1_2\,,\, \left(
   \begin{array}{cc} 0 & -e^i \\ e^i & 0 \end{array}\right) ,
   \left( \begin{array}{cc} 0 & 1\\ 1 & 0 \end{array}\right),
   \left( \begin{array}{cc} 1 & 0\\ 0 & -1 \end{array}\right)
   \right) \quad ,
\end{equation}
$k=0,i,4,5$. Making a similarity transformation $\sigma'^{\,k} = A
\sigma^k A^{-1}$ with $A=\frac{1}{\sqrt{2}}(\sigma^2+\sigma^3)$
($A^2= 1, A= A^{-1}$) and using the realization (\ref{neq4.7}) in
the expression of the $\sigma'^{\,k}$, we obtain six complex
$4\times 4$ matrices $\Sigma_{A\dot{B}}$ in the form
\begin{equation}
\label{eq6. 4}
 \left( 1_4,
\begin{pmatrix}-\sigma^i & 0 \cr 0 &\sigma^i
\end{pmatrix},
-\begin{pmatrix}0 & 1_2 \cr 1_2 & 0
\end{pmatrix}, i
\begin{pmatrix}0  & -1_2 \cr 1_2 & 0
\end{pmatrix}
\right)\ .
\end{equation}
Since $C\Sigma C^{-1}=\Sigma^*$, the undotted $A$ and the dotted
$\dot{A}$ indices transform similarly under $SO(1,5)$ and, unlike
in the $D=4$ case, there is no metric allowing us to raise and
lower the $D$=6 Weyl spinorial indices.

\indent
Let $\Lambda^{1 A} \equiv \Lambda^{A}$. It follows
that\footnote{Since the $\Sigma$'s are now four-dimensional and
$\tilde{\Sigma}$= $\Sigma^{\dot{B}A}\equiv(\Sigma^0, -\Sigma^s)$,
$s=1,\dots,5$, $\Sigma^k \widetilde{\Sigma}^l + \Sigma^l
\widetilde{\Sigma}^k =2\eta^{lk} 1_4$ ($\eta_{kl} = \hbox{diag}
(1, -1, \ldots -1)$) we define now $a_{A\dot{B}}\equiv \frac{1}{2}
\Sigma^k_{A\dot{B}}a_k$, $b^{\dot{B}A} \equiv
\frac{1}{2}\Sigma^{\dot{B}A\ k}b_k$. As a result, we have
$a_{A\dot{B}}b^{\dot{B}A}= a\cdot b$ as for $D$=4, but now
$\frac{1}{2}a_{A\dot{B}} \Sigma^{\dot{B}A\ k} = a^k$,
$\frac{1}{2}a^{\dot{B}A}\Sigma^k_{A\dot{B}} = a^k$, $k=\mu, 4, 5$.
The various $\sqrt{2}$'s in (\ref{eq6.5}) come from having
$D$=4-adapted factors in eqs. (\ref{eq1.2}), (\ref{eq1.4}) (see
footnote 1) within a $D$=6 context.}
\begin{eqnarray}
\label{eq6.5} \Lambda^{A}\; (\Sigma^{\mu})_{A\dot{B}} \;
\Lambda^{\dot{B}} &=& \lambda_{\alpha i}\; (\sigma^{\mu})^{\alpha
\dot{\beta}} \bar{\lambda}_{\dot{\beta}i}\;=\;\sqrt{2}p^\mu
 \ ,\cr
 \Lambda^{A} \;(\Sigma^{4})_{A\dot{B}} \; {\Lambda}^{\dot{B}} & = &
-\frac{1}{\sqrt{2}}(M+\bar{M})=\sqrt{2}p^4
 \ ,\cr
\Lambda^{A}\; (\Sigma^{5})_{A\dot{B}} \; {\Lambda}^{\dot{B}} &=&
-\frac{i}{\sqrt{2}}(M-\bar{M})= \sqrt{2}p^5 \; ,
\end{eqnarray}
where we have used eqs. (\ref{eq1.2}) ((\ref{eq1.4})) in the first
(second and third) expression above and that $M=p_4-ip_5\ ,\
\bar{M}=p_4+ip_5$. This gives the algebraic $D$=6 zero mass shell
condition,
\begin{equation}
\label{eq6.6}
p _k  p^k  = p_\mu p^\mu - p^2_4 - p^2 _5 = 0\; .
\end{equation}
Then, the bosonic lagrangian (\ref{eq2.2}) may be now written in a
six-dimensional form as the lagrangian for the $D$=6 massless
particle in a hybrid spinorial/spacetime formulation,
\begin{equation}
\label{eq6.7}
    {\cal L}^B = \sqrt{2}\Lambda^A \; \dot{x}_{A\dot{B}} \; \Lambda^{\dot{B}} \; ,
\end{equation}
where
\begin{equation}
\label{eq6.8}
    \dot{x}_{A\dot{B}} = \frac{1}{2} \dot{x}_k  (\Sigma^k)_{A\dot{B}}\;,
\end{equation}
and the $D$=6 zero mass condition \eqref{eq6.6} is built in
algebraically.

\indent   To supersymmetrize the bosonic lagrangian (\ref{eq6.7})
we introduce $D$=6 superspace with Weyl-Grassmann spinors. The
lagrangian (\ref{eq3.2}) is then obtained by the replacement
 \begin{equation}
 \label{eq6.9}
\dot{x}^{\dot{B}A} \longrightarrow \omega^{\dot{B}A} =
\dot{x}^{\dot{B}A} - 2i(\dot{\theta}^{A} \; \bar{\theta}^{\dot{B}}
- \theta^A \; \dot{\bar \theta}^{\dot{B}} )\, ,
\end{equation}
where we use the following four-component $D$=6 Grassmann spinors
\begin{equation}
\label{eq6.11}
\theta^A =
\left(
\begin{array}{c}
\theta^{\alpha } _{\ 1} \cr \bar{\theta}_{\dot{\alpha}  2}
\end{array}
\right) \ ,\qquad \theta^{\dot{A}} = \left(
\begin{array}{c}
\bar{\theta}^{\dot{\alpha} }_{\ 1} \cr {\theta}_{\alpha
 2}
\end{array}
\right) \; .
\end{equation}
The substitution (\ref{eq6.9}) may be now written in terms of a
pair of two-dimensional $D=4$ Weyl spinors as
\begin{eqnarray}\label{eq6.10}
&&\dot{x}_\mu \longrightarrow \omega_\mu = \dot{x}_\mu
 - i (\dot{\theta}^{\alpha}_{\ r} (\sigma_\mu )_{\alpha \beta}
 \; \bar{\theta}^{\dot{\beta}}_{\ r}
- \theta^{\alpha} _{\ r} (\sigma_\mu )_{\alpha \dot{\beta}} \;
\dot{\bar{\theta}} ^{\dot{\beta}}_{\ r} ) \ ,\cr
&&\dot{z}
\longrightarrow \omega = \dot{z} + 2i \; \theta_{\alpha r}\;
 \dot{\theta}^{\alpha r}\ ,\cr
 &&\dot{\bar{z}} \longrightarrow \bar{\omega} = \dot{{\bar{z}}} + 2i
\; \bar{\theta}_{\dot{\alpha} r} \dot{\bar{\theta}}^{\dot{\alpha
r}}\ .
\end{eqnarray}

\indent    It may be checked that the supersymmetric lagrangian
(\ref{eq3.2}) in $D=6$ notation can be written as follows
\begin{equation}\label{eq6.12}
{\cal L}^{SUSY} =\sqrt{2} \Lambda^A \; \omega_{A\dot{B}} \;
\Lambda^{\dot{B}} = \frac{1}{\sqrt{2}} \omega_k \Lambda^A
\Sigma^k_{A\dot{B}} \Lambda^{\dot{B}} \quad ,
\end{equation}
where $k=0,1,\dots4,5$ and
$\omega^k=(\,\omega^\mu\,,\,\frac{1}{2}(\bar{\omega}+\omega)\,,\,
\frac{i}{2}(\bar{\omega}-\omega)\,)$ and $\omega$ and
$\bar{\omega}$ are given in eq. (\ref{eq3.1b}).

\indent In our model (see (\ref{eq3.2}) or (\ref{eq6.12})) the
central charge coordinates $z$, $\bar{z}$, as well as the dual
central charges $M$, $\bar{M}$, are dynamical variables; however
the central charges are constants on-shell (the field equations
are $\dot{M}=\dot{\bar{M}}=0$). The static approximation
$M=\bar{M}={\textrm const.}$ can be achieved consistently in our
first order formulation by the $D=6\to D=4$ reduction procedure in
target space. We set in eq. (\ref{eq6.12}))\footnote{See
\cite{21azluk,22azluk}; for the application of the dimensional
reduction procedure to superparticles see
\cite{23azluk}-\cite{25azluk}. There, one performs the dimensional
reduction procedure in target space, in consistency with the
on-shell values of the reduced solutions. Note that there is no
need of restricting $x_4$, $x_5$ because in the first order
formalism the only term in the lagrangian depending on these
variables becomes a total derivative for constant $p_4 , p_5$.}
\begin{equation}
\label{eq6.13} p_4 = p_5 = \hbox{const.}
  \qquad \qquad p^2_4 + p^2_5= m^2 = \hbox{const.}
\end{equation}
Putting
\begin{equation}
\label{eq6.14}
    p_4 = m \sin \varphi \ ,\qquad \qquad p _5 = m  \cos \varphi\
    , \qquad \qquad M=me^{-i\varphi}\ ,
\end{equation}
where $\varphi$ is a constant phase, the dimensional reduction
${\cal L}^{SUSY} \to  {\cal L}^{SUSY}_{D=4}$ gives
\begin{equation}
\label{eq6.15}
    {\cal L}^{SUSY}_{D=4}
= p_\mu \omega^{\mu} + im \left( e^{-i\varphi } \; \theta_{\alpha
r} \; \dot{\theta}^{\alpha r} + e^{ i \varphi } \;
\bar{\theta}_{\alpha r} \; \dot{\bar{\theta}}^{\alpha r}  \right)+
\frac{e}{2} (p^2 - m^2) \; ,
\end{equation}
where the $D$=4 mass shell condition, eq. (\ref{eq6.6}) after
using (\ref{eq6.14}), is imposed by a Lagrange multiplier. We
further observe that we can set $\varphi =0$ because $\omega^\mu$
is invariant under the constant phase transformations
\begin{equation}
\label{eq6.16}
{ \theta}'_{\alpha r} = e^{\frac{i}{2}  \varphi } \;
\theta_{\alpha r} \quad , \quad { \bar{\theta}}'_{\dot{\alpha} r} =
e^{- \frac{i}{2} \varphi } \; \bar{\theta}_{\dot{\alpha} r}\quad .
\end{equation}
Subsequently, we obtain the first order formulation of the $D=4$,
$N=2$ superparticle model with WZ term introduced by the two of
present authors  \cite{17azluk}. Indeed, after eliminating $p_\mu$
and $e$ from (\ref{eq6.15}) by the algebraic field equations we
obtain
\begin{equation}
\label{eq6.17}
    {\cal L}^{SUSY}_{D=4}
= m \sqrt{\omega_\mu \; \omega^\mu}
 + i\; m (\theta_{\alpha r} \; \dot{\theta}^{\alpha r}
 +\bar{\theta}_{\alpha r} \;\dot{\bar{\theta}}^{\alpha r})\, .
\end{equation}

\indent The model (\ref{eq6.17}) corresponds to the case where the
central charge is represented by a constant real mass parameter.
It is worth stressing here that the equality of parameters $m$ in
front of the first (Nambu-Goto-like) and second (WZ) term in the
Lagrangian (\ref{eq6.17}) corresponds in our two-supertwistor
model to the equality of the numerical coefficients in front of
the two terms in (\ref{eq3.19}), necessary for the invariance
under the local $\kappa$-transformations \eqref{neq4.5} in
two-supertwistor space. In the $N=2$ superspace formulation, the
equality of the `bosonic' and `fermionic' masses in the two terms
of the lagrangian (\ref{eq6.17}) allows as well for the invariance
under the $\kappa$-gauge transformations \eqref{eq3.15x},
\eqref{eq3.16x}, which are necessary to balance the number of
fermionic and bosonic degrees of freedom in the $p$=0
super-$p$-brane model, as it is the case for extended objects in
general \cite{AETW87}.

\section{Discussion}
\setcounter{equation}{0}

 We have considered here the
supertwistorial formulation of $D$=4 superparticles with mass and
$N=2$ supersymmetry. Our interest in the massive case is due to
the fact that it is the massive superparticle model with WZ term
\cite{17azluk}, rather than the massless one, which is the
pointlike $p$=0 analogue of the extended $p>0$ super-$p$-branes.
By constructing a model with two $N=2$, $D=4$ supertwistors we
have been able to study the appearance of both the WZ term and the
fermionic $\kappa$-transformations in a (super)twistorial
framework.

\indent It is known that massless superparticles with $N$-extended
supersymmetry can be described by a single $N$-extended
supertwistor
 ${\cal Z}=(\lambda_{\alpha}, \bar{\omega}^{\dot{\alpha}},\eta_r)$
 with $N$ complex Grassmann
coordinates $\eta_r,\, r=1\dots N$. The degrees of freedom of one
superstwistor are invariant under $\kappa$-transformations {\it
i.e.}, for massless superparticles the coordinates of the single
supertwistor already describe the `physical' degrees of freedom.
Thus, since the $N$-extended $D=4$ superspace contains $2N$
complex Grassmann coordinates $\theta_{\alpha r}$ ($\alpha = 1,2$,
$r=1, \ldots N$), the equivalence between the supertwistorial and
the superspace formulations of the massless superparticle requires
the removal of half of the odd superspace degrees of freedom by
means of the fermionic gauge $\kappa$-transformations
\cite{17azluk,Sie83}.

\indent However, if we wish to describe a $D$=4 massive
superparticle in a supertwistorial approach, we necessarily need
at least two supertwistors to allow for a non-vanishing mass
\cite{Perjes,7azluk,3azluk},\cite{8azluk}-\cite{12azluk}. In this
paper we have considered the $N=2$ supersymmetry case using
 for two $D=4$ supertwistors, which give
rise to a supertwistorial WZ term and to the $\kappa$-gauge
transformations. Indeed, it turns out that the number of
Grassmannian degrees of freedom of our supertwistorial model is
the same as in $N$=2 superspace (see eqs. (\ref{eq3.17})), and
thus the familiar local fermionic transformations of the
superspace framework must appear as well in the purely
two-supertwistorial description.

\indent In order to obtain the $\kappa$-invariant formulation of
our model we observe that two $N=2$, $D=4$ supertwistors can be
obtained from a single $N=1$, $D=6$ supertwistor provided that the
odd $D=4$ supertwistor coordinates satisfy the $SU(2)$-Majorana
reality condition (eq. \eqref{neq4.3}). One can conclude therefore
that our $\kappa$-transformations account for the degrees of
freedom that disappear when we use such a pair of constrained
$D=4$ complex supertwistors, equivalent to the single $D=6$
supertwistor with associated quaternionic geometry.

 \indent
For the $D$=4, $N$-extended supersymmetry case one can introduce
$\frac{N(N-1)}{2}$ mass-like parameters corresponding to as many
complex central charges ($i=1 \ldots N>2$) \cite{26azluk}
\begin{eqnarray}
\label{eq7.1} \left\{ Q^i _{\alpha}, \bar{Q}^{j}_{\dot{\beta}}
\right\} &=& \delta^{ij} \; P_{\alpha \dot{\beta}} \quad, \cr
\left\{ Q^i _{\alpha}, Q^{j}_{\beta} \right\} = \epsilon_{\alpha
\beta} \; \Xi^{ij}  &,& \left\{ \bar{Q}^i _{\dot{\alpha}},
\bar{Q}^{j}_{\dot{\beta}} \right\} = \epsilon_{\dot{\alpha}
\dot{\beta}} \; \bar{\Xi}^{ij} \quad ,
\end{eqnarray}
where ${\Xi}^{ij} = -{\Xi}^{ji}$ is the complex $N\times N$
skewsymmetric matrix of generators of the central charges . To
introduce in a supertwistorial formalism all possible massive
parameters as independent spinorial bilinears, one can generalize
the relation $\Xi=M$ (eq. (\ref{Xivalue})) to allow for
antisymmetric charges as follows
\begin{equation}
\label{eq7.2}
    {\Xi}^{ij} = -{\Xi}^{ji} \propto
    \lambda^{i}_{\alpha} \; \lambda^{\alpha j} \; ,\quad i,j=1,\dots N \;
\end{equation}
({\it cf.} (\ref{eq1.4})). For such a purpose $N$ independent
copies of $N$-extended supertwistors (\ref{eq1.7}) are needed,
with $N^2$ complex fermionic degrees of freedom\footnote{Since for
$D$=4 there are two linearly independent constant spinors, the
$N>2$ case is useful for $D>4$ (the number of components of a
Dirac spinor grows as $2^{[\frac{D}{2}]}$).}. Superparticle models
characterized by having several mass-like parameters corresponding
to the central charges (\ref{eq7.2}) are not known, but by
generalizing of our two-supertwistor framework one may guess how
to construct a supertwistor lagrangian in terms of $N>2$ copies of
$N$-extended supertwistors. One of the primary tasks in building
such a model in $D$=4 would be to describe the corresponding
generalized $\kappa$-transformations which would require the
maximal number $N(N-1)$ of odd complex parameters.

\indent The most interesting case one could study is that of a
$D=4$, $N=4$ model with 12 $\kappa$-gauge odd parameters. If we
could introduce four $D=4$ supertwistors with suitable constraints
to describe the degrees of freedom of an octonionic $N=1$, $D=10$
supertwistor \cite{6azluk}, the $\kappa$-invariant formulation
would determine the corresponding $N=4,D=4$ supertwistor dynamics
with octonionic structure\footnote{In such a formalism a single
octonion coordinates spanning $\mathbb{R}^8$ would be described by
four complex split octonionic units (see {\it e.g.} \cite{GG76}).
It is unclear, however, whether the problems associated with
non-associativity can be avoided.}.

 \indent
We conclude by mentioning that, recently, there has been a renewed
interest in twistor theory and in the general Penrose programme as
a result of the applications of twistors and supertwistors in
various modern contexts as {\it e.g.},  in the analysis and
computation of $N$=4 Yang-Mills amplitudes
\cite{Witten03,CaSvWi04},  in various (super)string models
\cite{Witten03,NB04,Siegel04,IB+JdA+C=2006} or in connection with
an algebraic description of the BPS states in M-theory
\cite{27azluk}. One may assume, therefore, that the study of
dynamical multi-supertwistorial models is a useful step towards a
further application of  (super)twistorial ideas to fundamental
interactions formalisms.

\subsection*{Acknowledgments}

The authors would like to thank Dima Sorokin for valuable
discussions. This work has been partially supported by research
grants from the Spanish Ministry of Science and Innovation
(FIS2008-01980, FIS2005-03989) and EU FEDER funds, the Junta de
Castilla y Le\'on (VA013C05), the Polish Ministry of Science and
Higher Education (J.L., NN202318534) and the EU `Forces Universe'
network (MRTN-CT-2004-005104).  

\end{document}